\documentstyle[aps,preprint]{revtex}

\begin{document}

\input epsf
\newcommand{\infig}[2]{\begin{center}\mbox{ \epsfxsize #1
                       \epsfbox{#2}}\end{center}}

\newcommand{\be}{\begin{equation}}
\newcommand{\nn}{\nonumber}
\newcommand{\ee}{\end{equation}}
\newcommand{\bea}{\begin{eqnarray}}
\newcommand{\eea}{\end{eqnarray}}
\newcommand{\wee}[2]{\mbox{$\frac{#1}{#2}$}}   
\newcommand{\unit}[1]{\,\mbox{#1}}
\newcommand{\degree}{\mbox{$^{\circ}$}}
\newcommand{\ltish}{\raisebox{-0.4ex}{$\,\stackrel{<}{\scriptstyle\sim}$}}
\newcommand{\vs}{{\em vs\/}}
\newcommand{\bin}[2]{\left(\begin{array}{c} #1 \\ #2\end{array}\right)}
\newcommand{\pred}{^{\mbox{\small{pred}}}}
\newcommand{\retr}{^{\mbox{\small{retr}}}}
\newcommand{\p}{_{\mbox{\small{p}}}}
\newcommand{\m}{_{\mbox{\small{m}}}}
\newcommand{\rs}[1]{_{\mbox{\small{#1}}}}	
\newcommand{\ru}[1]{^{\mbox{\small{#1}}}}
\draft

\title{Retrodiction as a tool for micromaser field measurements}

\author{John Jeffers${}^1$, Stephen M. Barnett${}^1$ and David T. Pegg${}^2$}
\address{${}^1$ Department of Physics and Applied Physics,
University of Strathclyde, Glasgow G4 ONG, United Kingdom.\\ ${}^2$
Faculty of Science, Griffith University, Nathan, Brisbane, Queensland Q111,
Australia.}

\date{\today}
\maketitle

\begin{abstract}
We use retrodictive quantum theory to describe cavity field
measurements by successive atomic detections in the micromaser. We
calculate the state of the micromaser cavity field prior to detection
of sequences of atoms in either the excited or ground state, for atoms
which are initially prepared in the excited state. This provides the
POM elements which describe such sequences of measurements.
\end{abstract}

\vspace{2cm}

\noindent Corresponding Author: John Jeffers, Phone 0141 548 3213, Fax 0141 552 2891,
email john@phys.strath.ac.uk.

\vspace{2cm}




\section{Introduction}
The advent of quantum information theory, coupled with an increased
technological ability to produce and study exotic entangled states of
the optical field has stimulated a re-examination of the fundamental
principles of quantum mechanics. One example of this is the recent
resurgence of interest in retrodictive quantum theory, in which the
likelihood of past events is inferred from measurements of the
present.  This subject was, until relatively recently, a simple problem
associated with time-reversal in quantum theory \cite{oldret}. It has
now evolved into a powerful technique for calculating retrodictive
conditional probabilities in quantum theory, particularly when the
knowledge of the output state produced by a state preparation device is
not complete. The crucial step in this was the use of Bayes' theorem
\cite{bayes} coupled with normal predictive quantum theory. This allows
a retrodictive state to be defined such that retrodictive and
predictive conditional probabilities are consistent \cite{newret}. The
theory can be applied to both closed systems, in which the full
evolution can be followed, and to open ones, in which the system of
interest interacts with an unmeasured environment
\cite{atomic,ampatt,master,prev}.

The predictive state of a system at any time during the evolution is
the prepared state evolved forwards in time. The `collapse of the
wavefunction' occurs at the measurement device when this evolved state
is projected onto the measurement basis. The retrodictive state at any
time between preparation and measurement is the measured state evolved
backwards in time. At the preparation time this state collapses onto
any one of a set of possible initially prepared states. In open
systems, where the system must be described by a density operator, we
can define a retrodictive density operator which can be evolved
backwards in time according to a retrodictive master equation
\cite{master}. 

The micromaser has been useful tool for quantum optical and cavity QED
experiments for many years \cite{micreview}. In the basic device
highly-excited Rydberg atoms are passed through a high-Q microwave
cavity, one of whose longitudinal modes has a frequency which is close
to one of the atomic transition frequencies. The atoms interact with
the cavity field, and the whole device behaves to a very good
approximation like a two-level atom interacting with a single-mode
quantum optical field. The atoms can be prepared and measured in what
amounts to either the excited or ground state. In a further refinement
resonant microwave pulses can be applied to the atoms prior to entering
and after leaving the cavity.  This allows the atoms to be prepared or
measured in any desired superposition of the excited or ground states.
The micromaser operates in what has been termed the strong-coupling
regime of cavity quantum electrodynamics, in which a single photon in
the cavity is a strong enough field to cause multiple excitations and
de-excitations of the atom.

Usually there is no direct way to measure the cavity field in the
micromaser. The only way to tell whether or not the cavity field has
been excited by an atom is to send a further atom into the cavity in a
known state and to measure the atomic state after it has passed through
the cavity. It would therefore be useful to know precisely what
measurement an atomic detection in a particular state (or a sequence of
detections in various states) makes on the cavity field. This paper is
designed to solve this problem for the case that all of the atoms enter
the cavity in the excited state. If an atom or sequence of atoms in a
known state are passed through the cavity and detected as a sequence of
atoms at the output, what does this particular sequence of detections
say about the cavity field prior to the passage of the atoms?

This paper is organised as follows. In Section (II) we give a brief
overview of retrodictive quantum theory. Section (III) describes how
this may be applied to the two-level atom interacting with a single
field mode, with particular emphasis on retrodicting micromaser field
states form atomic state measurements.  Section (IV) discusses the
results of particular measurements and in Section (V) contains a
summary and discussion of the results.

\section{Quantum Retrodiction} 
In this section we provide brief details of retrodictive quantum
theory. A fuller account can be found in references
\cite{newret,atomic,ampatt,master,prev}. Elsewhere in this volume we
provide a more complete discussion of the quantum theory of preparation
and measurement \cite{prepmeas}. The analysis which follows, based on
preparation operators and positive operator measures (POMs) \cite{pom}
is necessarily incomplete, but sufficient for our purposes.

Suppose that we have a preparation device which produces output states
$\hat{\rho}\pred_i$ with prior probabilities $P(i)$.
$\hat{\rho}\pred_i$ is the usual density operator of predictive quantum
mechanics. This state can evolve and interact with other systems until
it is measured by a measuring device. A measurement in which none of
the measurement results are lost or discarded can be described by a
measurement POM \cite{pom}. This is a set of non-negative definite,
Hermitian elements $\hat{\Pi}_j$ which sum to the unit operator, each
element corresponding to a particular measurement outcome. In general
there is no requirement that there be the same number of POM elements
as there are states which span the system space, but for von Neumann
measurement this is so, and the POM elements are simply the projectors
of the particular chosen states which span the space.  Suppose that
preparation takes place at time $t\p$ and measurement at a later time
$t\m$. Within this framework the predictive probability that the
measurement outcome $\hat{\Pi}_j$ is obtained given that the state
$\hat{\rho}\pred_i$ was prepared is
\bea
\label{predconprob}
P(j|i)=\mbox{Tr}\left(\hat{\rho}\pred_i (t\m) \hat{\Pi}_j\right),
\eea
where 
\bea
\label{rhopredclosed}
\hat{\rho}\pred_i(t\m)= \hat{U}(\tau)
\hat{\rho}\pred_i(t\p) \hat{U}^\dagger(\tau)
\eea is the initial density operator evolved forwards in time to the
collapse time and
\bea
\label{evolve}
\hat{U}(\tau) = \exp \left(-\frac{i}{\hbar} \hat{H} \tau\right)
\eea
is the evolution operator, which operates for the length of time
between preparation and measurement, $\tau = t\m-t\p$.

Suppose that instead of calculating the predictive probability $P(j|i)$
we wish to calculate the retrodictive conditional probability $P(i|j)$
that the state $\hat{\rho}\pred_i$ was prepared, given a particular
measurement result $\hat{\Pi}_j$. It is possible to do this by
calculating all possible predictive conditional probabilities for the
system, and then using Bayes' theorem. A simpler and more natural
approach is to use retrodictive quantum theory, so that the required
probability can be written \cite{newret,atomic}
\bea
\label{retconprob}
P(i|j)= \frac {\mbox{Tr} \left[ \hat{\Lambda}_i \hat{\rho}\retr_j
(t\p) \right]} {\mbox{Tr} \left[ \hat{\Lambda}
\hat{\rho}\retr_j (t\p) \right]}.
\eea Here the operator $\hat{\Lambda}_i$ is the preparation device
operator, and
\bea
\hat{\Lambda}=\sum_i \hat{\Lambda}_i = \sum_i P(i)
\hat{\rho}\pred_i,
\eea is the {\it a priori} density operator, the sum of each possible
preparation density operator weighted by its prior probability of
production.  $\hat{\Lambda}$ is the best description of the state we
can give without knowing the outcome of the preparation or
measurement.  The retrodictive density operator at the preparation time
is simply the normalised measurement POM element evolved back from the
measurement time to the preparation time,
\bea
\label{rhoretrclosed}
\hat{\rho}\retr_j (t\p) = \hat{U}^\dagger(\tau)
\hat{\rho}\retr_j(t\m) \hat{U}(\tau),
\eea with
\bea
\hat{\rho}\retr_j(t\m) =
\frac{\hat{\Pi}_j}{\mbox{Tr}\hat{\Pi}_j}.
\eea

The above formulae for the conditional probabilities (eqs.
(\ref{predconprob}) and (\ref{retconprob})) apply equally well for open
systems, where the system of interest interacts with an unmeasured
environment with many degrees of freedom. If this environment causes
information to be lost about the system, and the Born-Markov
approximation holds, then the the evolved density operators are the
solutions of master equations \cite{jch}. In eq.(\ref{predconprob}) the
density operator required for such an open system, which for a closed
system would be given by eq. (\ref{rhopredclosed}), is the solution of
the usual master equation forwards in time from the preparation  time
to the measurement time.  However, in eq. (\ref{retconprob}) the
solution required instead of eq.  (\ref{rhoretrclosed}) is that of the
retrodictive master equation, giving the evolution backwards in time
from the measurement time to the preparation time. We have recently
derived this equation from the general principle that conditional
probabilities should be independent of the time of collapse of the
wavefunction \cite{master}.  This principle, which applies equally well
in open or closed systems, leads to an unusual interpretation of
measurements in a quantum system. The measurement POM elements, when
evolved back in time, still satisfy the criterion that they form a
valid POM, namely that they sum to the unit operator. This means that
the evolved POM still represents a valid measurement of the quantum
system, and we may regard any amount of the evolution as still forming
part of the measurement. The same may be said of the prepared density
operator evolved forwards in time. 

In the system considered in the present
paper, however, the Born-Markov approximation is not made, and so
we must consider the full evolution of the coupled atom-field system.


\section{Prediction and retrodiction for the coupled atom-field 
system }
\label{sec:predret}

Here we apply the retrodictive formalism to a coupled system consisting
of a two-level atom and a single cavity mode of an electromagnetic
field. We have already applied retrodictive quantum mechanics to this
system.  When the field is excited by a coherent state, predictively
there are collapses and revivals of the Rabi oscillations in the atomic
excitation probability.  If the atom is measured to be in a particular
state there also exist earlier revivals or `previvals' in the atomic
excitation at times prior to the measurement \cite{prev}. Here we will
concentrate specifically on the micromaser, in which atoms pass through
a microwave cavity one at a time. The interaction between an atom with
upper level $|e\rangle$ and lower level $|g \rangle$, and an
electromagnetic field is governed by the Jaynes-Cummings Hamiltonian
\cite{jch}. In the interaction picture this is
\bea
\label{jcham}
\hat{H}= \frac{\hbar \Delta}{2} \hat{\sigma}_3-i \hbar \lambda
\left(\hat{\sigma}_+\hat{a}-\hat{a}^\dagger \hat{\sigma}_- \right),
\eea 
where $\Delta$ is the detuning between the atomic frequency and the
cavity mode, $\hat{\sigma}_3=|e\rangle \langle e|-|g\rangle \langle g|$
is the atomic inversion operator, $\hat{\sigma}_+ = |e\rangle \langle
g|$ and $\hat{\sigma}_-= |g\rangle \langle e|$ are the atomic raising
and lowering operators, $\hat{a}^\dagger$ and $\hat{a}$ are the
creation and annihilation operators for the single mode field, and
$\lambda$ is the coupling constant. The rotating wave approximation,
which has been made in deriving this Hamiltonian, ensures that whenever
a photon is lost from the field the atomic state must change from
$|g\rangle$ to $|e\rangle$. In the standard predictive picture of
quantum mechanics a coupled atom-field system evolves forwards in time
according to this Hamiltonian from a preparation time $t\rs{p}$ to a
measurement time $t\rs{m}$.  After this has happened the coupled
density operator for the whole system is
\bea
\label{coupred}
\hat{\rho}\pred \rs{af}(t\rs{m}) = \hat{U}(\tau) \hat{\rho}\pred
\rs{a}(t\p) \otimes \hat{\rho}\pred \rs{f}(t\p) \hat{U}^\dagger(\tau),
\eea 
where $\hat{\rho}\pred \rs{af}(t\m)$ is the coupled density operator
for the atom-field system at the measurement time, $\hat{\rho}\pred
\rs{a}(t\p)$ and $\hat{\rho}\pred \rs{f}(t\p)$ are the uncoupled atom
and field density operators at the preparation time. $\hat{U}(\tau)$ is
the evolution operator for the coupled system, given by
eq.(\ref{evolve}) with Hamiltonian given by eq. (\ref{jcham}).  In the
micromaser we generally wish to infer the state of the cavity field
from a measurement of the atom. This is done by conditioning on the
atomic measurement outcome and tracing over the atomic states,
\bea
\label{rhofpred}
\hat{\rho}\pred\rs{f}(t\m) \propto
\mbox{Tr}\rs{a}\left[\hat{\Pi}\rs{a}\hat{U}(\tau) \hat{\rho}\pred \rs{a}(t\p)
\otimes \hat{\rho}\pred \rs{f}(t\p) \hat{U}^\dagger(\tau)\right],
\eea
where $\hat{\Pi}\rs{a}$ is the POM element corresponding to the outcome
of the measurement performed on the atom. If no measurement is
performed on the atom then $\hat{\Pi}\rs{a}=\hat{1}\rs{a}$. Similarly
we may not have any information about the initial state of the field,
in which case $\hat{\rho}\pred \rs{f}(t\p)\propto \hat{1}\rs{f}$.

The retrodictive picture differs from above in that the state of the
system before the measurement is assigned on the basis of the
measurement outcome. Thus if the measurement POM elements for the atom
and the field are $\hat{\Pi}\rs{a}(t\m)$ and $\hat{\Pi}\rs{f}(t\m)$,
the coupled initial density operator corresponding to equation
(\ref{coupred}) is
\bea
\label{coupret}
\nonumber \hat{\rho}\retr \rs{af}(t\p) &=& \hat{U}^\dagger(\tau)
\hat{\rho}\retr \rs{a}(t\m) \otimes \hat{\rho}\retr \rs{f}(t\m)
\hat{U}(\tau)\\
&\propto& \hat{U}^\dagger(\tau) \hat{\Pi}\rs{a}(t\m)
\otimes \hat{\Pi}\rs{f}(t\m) \hat{U}(\tau).
\eea 
In the micromaser the atom will in general have been prepared in a
particular initial state, and this state conditions the coupled density
operator similarly to equation (\ref{rhofpred}), to give the
retrodictive density operator for the field
\bea
\label{rhofretr}
\hat{\rho}\retr \rs{f}(t\p) \propto \mbox{Tr}\rs{a}
\left[\hat{\rho} \pred \rs{a}(t\p) \hat{U}^\dagger(\tau) \hat{\Pi}\rs{a}(t\m)
\otimes \hat{\Pi}\rs{f}(t\m) \hat{U}(\tau)\right],
\eea
where the constant of proportionality is determined by normalisation.
If there is no information at all about the preparation of the initial
states then $\hat{\rho}\pred \rs{a}(t\p)$ becomes proportional to the 
unit operator for the atomic space. Equation (\ref{rhofretr}) is enough to 
provide all of the information needed for this paper.


\section{Retrodiction of micromaser field states}
\label{sec:micromaser}
In this section we apply the formalism derived in the previous section
to the micromaser. Two-level atoms in their excited state are passed
through an optical cavity and a measurement of their state is
performed. We assume that the atomic frequency is resonant with the
micromaser field mode. This is a typical set-up of a micromaser
experiment \cite{micreview,fock}.  The final field state is unmeasured
and we wish to know what the initial field state was before the atoms
passed through. The formalism can equally well be applied for any other
initial atomic state, but we will leave this for a further publication.
In order to find the initial field state from a sequence of atomic
measurements we first consider the effect of a single atom passing
through the cavity.  Equation (\ref{rhofretr}) provides the conditioned
retrodictive field state given knowledge of the initial and final
atomic states. The preparation time $t\rs{p}$ is effectively the time
at which the atom enters the cavity, and as the atomic state does not
change after the atom leaves the cavity, the exit time is the
measurement time $t\rs{m}$.

Before their passage through the cavity the atoms are initially in
their excited state, so $\hat{\rho}\rs{a}(t\rs{p})= |e\rangle \langle e|$.
The trace over the atomic states is trivial, and the initial field
state is
\bea
\hat{\rho} \retr \rs{f}(t\rs{p}) &\propto& \langle e|
\hat{U}^\dagger(\tau) \hat{\Pi}_{a}(t\rs{m}) \otimes \hat{\Pi}_{f}(t\rs{m})
\hat{U}(\tau) |e\rangle.
\eea 
We assume that the cavity POM element at $t\m$ is given 
by 
\bea
\Pi \rs{f}(t\rs{m}) \propto \hat{\rho} \rs{f}(t\rs{m}) = 
\sum_n{P_n(t\rs{m}) |n\rangle \langle n|}, 
\eea 
where $P_n(t\rs{m})$ is the probability that the field contains $n$
photons at $(t\rs{m})$.  If the atom is detected in the excited state
then the POM element (and retrodictive density operator) is simply the
excited state projector.  The unitary evolution operator is given by
equation (\ref{evolve}). A little algebra gives the result
\bea
\hat{\rho} \retr \rs{f}(t\rs{p})  \propto \sum_n{P_n(t\rs{m})
\cos^2[\Omega(n+1) \tau /2]|n\rangle \langle n|},
\eea
where $\Omega(n) = 2\lambda \sqrt{n}$. Thus the initial state photon
number probabilities are equal to the final state probabilities
multiplied by a cosine function whose argument is the product of the
Rabi frequency and the interaction time. If the atom is
detected in the ground state a similar calculation gives
\bea
\hat{\rho} \retr \rs{f}(t\rs{p})  \propto \sum_n{P_{n+1}(t\rs{m}) 
\sin^2[\Omega(n+1) \tau /2]|n\rangle \langle n|}.
\eea
In this case the probability that the initial state contains $n$ photons
depends upon the final state probability for $n+1$ photons and a sine
function which contains the $n+1$-photon Rabi frequency. The occurrence
of $P_{n+1}(t\rs{m})$ in the above formula is easy to explain, as when
the atom exits in the ground state there is one more photon in the
final field than the initial.

Successive atoms entering the cavity can be treated one at a time. If two
atoms enter the cavity in sequence then the initial retrodictive state into which the
cavity is projected by the detection of the second atom becomes the
final state in an identical calculation to the above for the first
atom entering the cavity. Thus every time an atom enters the cavity and
leaves in its excited state the photon number probabilities undergo the
transformation
\bea
P_n(t\rs{p}) \propto P_n(t\rs{m})  \cos^2[\Omega(n+1) \tau /2],
\eea
and when an atom leaves in the ground state the transition is 
\bea
P_n(t\rs{p}) \propto P_{n+1}(t\rs{m})  \sin^2[\Omega(n+1) \tau /2].
\eea 
The exact retrodictive density operator can be found by normalisation
at the end of the calculation by simply requiring all of the
probabilities at $t\rs{p}$ for the first atom to sum to unity.

As a simple example consider two atoms passing through the cavity.
There are four possible pairs of detection events which are summarised
in the table below.
\begin{center}
\vskip 0.5cm
\begin{tabular}{|l|l|}\hline
detection events & initial photon number relative probability \\ \hline\hline
$|e \rangle$ then $|e \rangle$ &  $P_n  \cos^4[\Omega(n+1) \tau/2]$ \\ \hline
$|g \rangle$ then $|g \rangle$ &  $P_{n+2}  \sin^2[\Omega(n+2) \tau/2]
\sin^2[\Omega(n+1) \tau/2]$ \\ \hline
$|e \rangle$ then $|g \rangle$ &  $P_{n+1}  \cos^2[\Omega(n+1) \tau/2]
\sin^2[\Omega(n+1) \tau/2]$ \\ \hline
$|g \rangle$ then $|e \rangle$ &  $P_{n+1}  \cos^2[\Omega(n+2) \tau/2]
\sin^2[\Omega(n+1) \tau/2]$ \\ \hline
\end{tabular}
\end{center}
{\small Table 1. Relative probability that the cavity contains n
photons prior to the passage and detection of two atoms.  $P_n$ is the
final state probability for $n$ photons after both atoms have been
detected.}
\vskip 0.5cm
It is clear from the last two lines of the table that the order of
detection matters. Every time an atom is detected in its ground state
the number of photons in the cavity increases, and the Rabi frequencies
increases correspondingly. Whether this occurs before or after the
detection of an atom in the excited state will affect the Rabi factor
imposed on the initial state probability.

Measurement outcomes are characterised by their corresponding POM
elements \cite{pom}.  The atom detection experiments make a measurement
of the cavity field, so there must be a POM element corresponding to
this measurement.  Different numbers and orderings of atomic detections
in either the excited or ground states correspond to different POM
elements \cite{pom}. For $s$ atomic detections in either the excited or
ground state there are a possible $2^s$ POM elements
$\hat{\Pi}_{1..s}$. These POM elements can be used to calculate the
probability for any given sequence of atom measurements given a known
initial field state density operator $\hat{\rho}\rs{f}(t\p)$. The POM
elements may be found by using an unmeasured state as the final cavity
state that, in the usual predictive formalism, would have the unit
operator for the cavity state space as its associated POM element. This
implies no knowledge of the final cavity state. It is convenient to set
an upper limit $N$ on the allowed photon number distribution. This
allows us to determine a well-behaved retrodictive field state. A
derivation of the field POM elements without this cut-off is given in
the appendix. Thus in place of the unit operator we can use the
operator
\bea
\hat{1}\rs{f} = \sum_{n=0}^{N} |n\rangle \langle n|
\eea 
corresponding to a final state $\hat{1}\rs{f}/(N+1)$. This is the unit
operator acting on the $N+1$-dimensional subspace spanned by the photon
number states up to and including $|N\rangle$. Backward time evolution
including the passage through the cavity of each of the $s$ atoms
yields the POM elements $\hat{\Pi}_{1..s}$ in the form
\bea
\hat{\Pi}_{1..s} = \sum_{n=0}^{N} C_{n,1..s} |n\rangle \langle n|
\eea
where the coefficients $C_{n,1..s}$ will be proportional to the
corresponding coefficients in the expression for $\hat{\rho} \retr
\rs{f}(t\rs{p})$. The POM elements must be normalised to sum to the
unit operator or, in this case, the unit operator $\hat{1}\rs{f}$ that
acts on the restricted photon subspace. For the two-atom case the
coefficients $C_{n,1..s}$ for the four POM elements are just the
trigonometric factors given in Table 1. For example $C_{n,ee}$ for the
detection of $|e\rangle$ then $|e\rangle$ is
$\cos^4[\Omega(n+1)\tau/2]$.

Figures 1-4 show photon number probabilities (and thus give the POM
elements) corresponding to particular sequences of atomic detections.
In each figure the interaction strength and time have been set so that
$\lambda \tau = \pi$.

Figure 1 shows the relative photon number probabilities for a sequence
of atoms detected in the excited state. If a single atom passes through
the cavity, and is detected the initial photon number distribution of
the cavity has the cosinusoidal form shown in figure 1(a). The
distribution has maxima at photon numbers which are one less than
perfect squares, corresponding to Rabi phase shifts of integral
multiples of $\pi$. If the cavity contains one of these numbers of
photons then an atom in the excited state must exit in the excited
state. This is what is known as the trapping state condition, and it
has been used to prepare photon number states of the electromagnetic
field \cite{fock}.  If a whole series of atoms pass through and are
detected in the excited state the troughs in the distribution become
wider as the passage of one more atom means that each photon number
probability gets multiplied by a cosinusoidal factor smaller than unity
for all photon numbers except those one less than perfect squares. This
is shown for five atoms in figure 1(b).

Detecting atoms in the ground state gives rise to a different cavity
POM element. In this case the minima of the sinusoidal distribution
fall at photon numbers which are one less than perfect squares. One of
the photons in the cavity when the atom is measured had to come from
atomic de-excitation so there had to be one fewer photon in the cavity
prior to the passage of the atom. Thus the whole distribution at $t
\rs{m}$ is shifted by one photon at $t \rs{p}$. The distribution for
one atom detected in the ground state is shown in figure 2(a).

Detecting atoms in the ground state thus has two effects:  (1) it
shifts the photon number probability distribution downwards by one
photon for each atom detected and, (2) for a system which satisfies the
trapping state condition it introduces zeros into the distribution .
The zeros occur because the Rabi phase shifts for excitation numbers
one greater than these photon numbers are $\pi$, and so atoms prepared
in the excited state can not exit in the ground state. These two
effects combine to allow the measurement of photon number distributions
with large gaps in them. For example, if three atoms are detected in
the ground state then the cavity field cannot have contained fewer than
four photons, as shown by figure 2(b). This POM element also has a
photon number gap from 6-8 photons, and there are also gaps at higher
photon number. These gaps extend their range downwards by one photon
with each extra ground state detection. This is shown in figure 2(c),
which is the photon number distribution prior to six atoms detected in
the ground state. The initial cavity photon distribution must have
contained at least nine photons. This also gives information about the
number of photons in the cavity after the detection. There must be at
least 15. Six of these come from the passing atoms and at least nine
were in the cavity initially.

If we have a little extra prior information about the cavity photon
number distribution this measurement can allow us to measure and
prepare specific photon number states. For example, say two atoms are
sent through the cavity and detected in the ground state. If we know
that there are not more than five photons in the cavity after the atoms
have passed through (not more than three prior to the atomic passage)
then these two atoms are sufficient to give an initial cavity
distribution of one photon for certain (figure 3).  This measures the
one photon number state. Furthermore, the fact that the cavity
initially contained one photon means that the de-excitation of the two
atoms has now created a three photon state. Other possible measurements
and preparations are possible, with the restriction on photon number a
little higher.

In general a set of atoms passing through the cavity will not all be
detected in the excited or ground state. Detection of any atoms in the
ground state will introduce zeros into the photon number distribution,
and these will move towards lower photon numbers with each successive
atom detected in the ground state.  Otherwise the distribution will be
peaked around photon numbers where the product of the cosinusoidal and
sinusoidal distributions is maximised. Examination of figures 1(a) and
2(a) shows that this will be around five and ten photons. This is borne
out by figure 4 which shows two typical results for six detected atoms,
alternating between detection in the excited and ground states. The
only significant probabilities are for 4-5 and 9-12 photons. Other
orderings give slightly different ratios of probability between these
numbers, but the results are similar. 


\section{Conclusions}
\label{sec:concs}

In this paper we have analysed field measurements in the micromaser.
The field is not directly measurable, but knowledge of the prepared
initial atomic state and the measured state after the interaction
between the atoms and the cavity gives information about the initial
cavity photon number distribution. If nothing is known {\it  a priori}
about the cavity field this allows us to calculate the POM elements for
the field corresponding to particular sequences of atomic detections in
either the excited or ground states.

We have considered the case where the atoms are all initially prepared
in the excited state. If an atom is detected in the excited state the
calculated initial photon number distribution is simply the
distribution after the measurement multiplied by a cosinusoidal Rabi
factor which depends upon one more than the final photon number. This
happens for each repeated detection in the excited state.  Atoms
detected in the ground state give the initial field a sinusoidal
distribution which depends upon the one more than the final photon
number.

If the atoms are measured in the excited state the cavity is much more
likely to have been in a photon number state corresponding to a
trapping state, for which the Rabi phase shift is a multiple of $\pi$.
Conversely, if the atom is measured in the ground state the cavity
field cannot have been in a trapping state. The probability of states
with these photon numbers is zero. Repeated measurement in the ground
state therefore measures photon number distributions with large gaps.
This can be used both to measure and prepare photon number states,
particularly where some information about the field also exists.

Earlier work on retrodictive master equations \cite{master}, and on the
derivation of the POM elements corresponding to homodyne detection (the
line states) \cite{ojthesis}, has shown that the boundaries between
preparation, evolution and measurement in quantum theory are arbitrary
choices. This is emphasised in the micromaser where the measurement
{\it and} preparation of particular atomic states corresponds to a
single measurement of the field. A measurement of a quantum system
necessarily includes all of the available information about the
system.  If any of this information should change so does the
measurement POM. If the atoms had been prepared in a superposition of
the excited and ground states then measurement in the excited or ground
state would correspond to different field POM elements to those
described here. We will explore preparation and measurement of the atom
in superpositions in further work.


\acknowledgments

We would like to thank Ben Varcoe for suggesting to us the application
of retrodiction to this problem. We also thank the United Kingdom
Engineering and Physical Sciences Research Council and the Australian
Research Council for financial support. SMB thanks the Royal Society of
Edinburgh and the Scottish Executive Education and Lifelong Learning
Department for the award of a Support Research Fellowship. JJ thanks 
Giovanna Morigi for useful discussions.

\renewcommand{\theequation}{A.\arabic{equation}}
\setcounter{equation}{0}
\section*{Appendix}
The sequence of detections of atoms in their ground or excited states
represents a measurement of the initial field. Our analysis has
concentrated on the retrodictive field state prior to interaction with
the first atom. It is not necessary, however, to introduce the
retrodictive state in order to analyse the field measurement. It can be
calculated simply by normalising the derived POM element. In
calculating the POM elements we have placed a limit on the photon
number of the final cavity field. This is also not strictly necessary
provided that we are only interested in the field measurement and its
associated POM. If we know the initial field state, $\hat{\rho} \pred
\rs{f} (t\p)$, then these POM elements give us the probabilities for
any given sequence of atom detections.

In constructing the POM, we start with the unmeasured field state after
the final ($s\ru{th}$) atom has passed throught the cavity. We
associate with this the trivial POM element
\bea
\hat{\Pi}\rs{f}\ru{final} = \hat{1}\rs{f} = \sum_{n=0}^{\infty} 
|n\rangle \langle n|.
\eea 
The POM element associated with immediately prior to detection of the
$s\ru{th}$ atom (denoted $\hat{\Pi}\rs{f}\ru{s}$) will depend upon
whether the atom was detected in the excited or ground state:
\bea
\nn \hat{\Pi}\rs{f}\ru{s}(e) &=& \langle e | \hat{U}^\dagger |e \rangle 
\langle e | \otimes \hat{1}\rs{f} \hat{U} |e \rangle\\
\nn &=& \cos \left[ \Omega(\hat{n}+1)\tau/2 \right] \hat{1}\rs{f} 
\cos \left[ \Omega(\hat{n}+1)\tau/2 \right]\\
&=& \cos^2 \left[ \Omega(\hat{n}+1)\tau/2 \right],
\eea
\bea
\nn \hat{\Pi}\rs{f}\ru{s}(g) &=& \langle e | \hat{U}^\dagger |g \rangle 
\langle g | \otimes \hat{1}\rs{f} \hat{U} |e \rangle\\
\nn &=& \sin \left[ \Omega(\hat{n}+1)\tau/2 \right] \hat{E} \hat{1}\rs{f} 
\hat{E}^\dagger \sin \left[ \Omega(\hat{n}+1)\tau/2 \right]\\
&=& \sin^2 \left[ \Omega(\hat{n}+1)\tau/2 \right].
\eea
Here $\hat{n}$ is the photon number operator and $\hat{E}(\hat{E}^\dagger)$ is
the bare lowering (raising) operator \cite{sussk}, 
\bea
\hat{E} &=& \sum_{n=0}^\infty |n \rangle \langle n+1| \\
\hat{E}^\dagger &=& \sum_{n=0}^\infty |n+1 \rangle \langle n|.
\eea
These operators have the properties
\bea
\hat{E}f(\hat{n}) &=& f(\hat{n}+1)\\
f(\hat{n})  &=& \hat{E}^\dagger f(\hat{n}+1)\\
\hat{E}\hat{E}^\dagger =\hat{1}\\
\hat{E}^\dagger\hat{E}= \hat{1}-|0 \rangle \langle 0|.
\eea
We can use these operators to write the POM elements associated with
any given sequence of atom detections by induction. This means starting
with the last atom detected and writing the POM element $\cos^2 \left[
\Omega(\hat{n}+1)\tau/2 \right]$ if the atom was in its excited state
and $\sin^2 \left[ \Omega(\hat{n}+1)\tau/2 \right]$ if it was in its
ground state.  For each preceding detection we pre- and postmultiply
the POM element by $\cos \left[ \Omega(\hat{n}+1)\tau/2 \right]$ if the
atom was measured in its excited state.  If the atom was found in its
ground state then we premultiply by $\sin \left[ \Omega(\hat{n}+1)
\tau/2 \right] \hat{E}$ and postmultiply by $\hat{E}^\dagger \sin
\left[ \Omega(\hat{n}+1)\tau/2 \right]$.

As an example, we consider a sequence of three atoms. The corresponding POM
elements $\hat{\Pi}\rs{f}(i, j, k)$ ($i, j, k = e, g$) are associated with the
atoms interacting with the cavity in the order $i, j$ then $k$. The eight
possible POM elements are.
\bea
\label{eee}
\hat{\Pi}\rs{f}(e, e, e) &=& \hat{c}^6(1),\\
\label{eeg}
\hat{\Pi}\rs{f}(e, e, g) &=& \hat{c}^2(1)\hat{s}(1) \hat{E} 
\hat{E}^\dagger \hat{s}(1) \hat{c}^2(1) = \hat{c}^4(1)\hat{s}^2(1),\\
\label{ege}
\hat{\Pi}\rs{f}(e, g, e) &=& \hat{c}(1) \hat{s}(1) \hat{E} \hat{c}^2(1)
\hat{E}^\dagger \hat{s}(1) \hat{c}(1) = \hat{c}^2(1) \hat{s}^2(1)
\hat{c}^2(2),\\
\label{egg}
\hat{\Pi}\rs{f}(e, g, g) &=& \hat{c}(1) \hat{s}(1) \hat{E} \hat{s}(1)
\hat{E} \hat{E}^\dagger \hat{s}(1) \hat{E}^\dagger \hat{s}(1)
\hat{c}(1) = \hat{c}^2(1) \hat{s}^2(1) \hat{s}^2(2),\\
\label{gee}
\hat{\Pi}\rs{f}(g, e, e) &=& \hat{s}(1) \hat{E} \hat{c}^4(1)
\hat{E}^\dagger \hat{s}(1)
= \hat{s}^2(1) \hat{c}^4(2),\\
\label{geg}
\hat{\Pi}\rs{f}(g, e, g) &=& \hat{s}(1) \hat{E} \hat{c}(1) \hat{s}(1)
\hat{E} \hat{E}^\dagger \hat{s}(1) \hat{c}(1) \hat{E}^\dagger
\hat{s}(1) = \hat{s}^2(1) \hat{c}^2(2) \hat{s}^2(2),\\
\label{gge}
\hat{\Pi}\rs{f}(g, g, e) &=& \hat{s}(1) \hat{E} \hat{s}(1) \hat{E} 
\hat{c}^2(1) \hat{E}^\dagger \hat{s}(1) \hat{E}^\dagger
\hat{s}(1) = \hat{s}^2(1) \hat{s}^2(2) \hat{c}^2(3),\\
\label{ggg}
\hat{\Pi}\rs{f}(g, g, g) &=& \hat{s}(1) \hat{E} \hat{s}(1) \hat{E}
\hat{s}(1) \hat{E} \hat{E}^\dagger \hat{s}(1) \hat{E}^\dagger
\hat{s}(1) \hat{E}^\dagger \hat{s}(1) = \hat{s}^2(1) \hat{s}^2(2)
\hat{s}^2(3).
\eea
Here we have written 
\bea
\hat{c}(m) &=& \cos \left[ \Omega(\hat{n}+m) \tau/2 \right],\\
\hat{s}(m) &=& \sin \left[ \Omega(\hat{n}+m) \tau/2 \right].
\eea

We can also summarise the construction of the field POM elements by iteration
starting from the first atom to enter the cavity:\\
(1) Start with the first atom to interact with the cavity.\\
(2) If the atom is measured in the excited state then write $\hat{c}^2(1)$. If
the atom is measured in the ground state them write $\hat{s}^2(1)$ and add 1 to
the arguments of all subsequent $\hat{c}$ and $\hat{s}$ operators.\\
(3) Follow the same prescription for each subsequent atom in turn.

\newpage

Figure Captions\\
Figure 1. Measured cavity photon number probability distributions for
atoms detected in the excited state (a) one atom, (b) five atoms.\\
Figure 2. Measured cavity photon number distributions for atoms detected in the
ground state (a) one atom, (b) three atoms, (c) six atoms.\\
Figure 3. Measured cavity photon number probability distributions for two
atoms detected in the ground state given that there were no more than three
photons in the cavity prior to the passage of the atoms.\\
Figure 4. Measured cavity photon number probability distributions for six
detected atoms, three in the ground state, three in the excited state in the
order (a) g, e, g, e, g, e and (b) e, g, e, g, e, g.\\

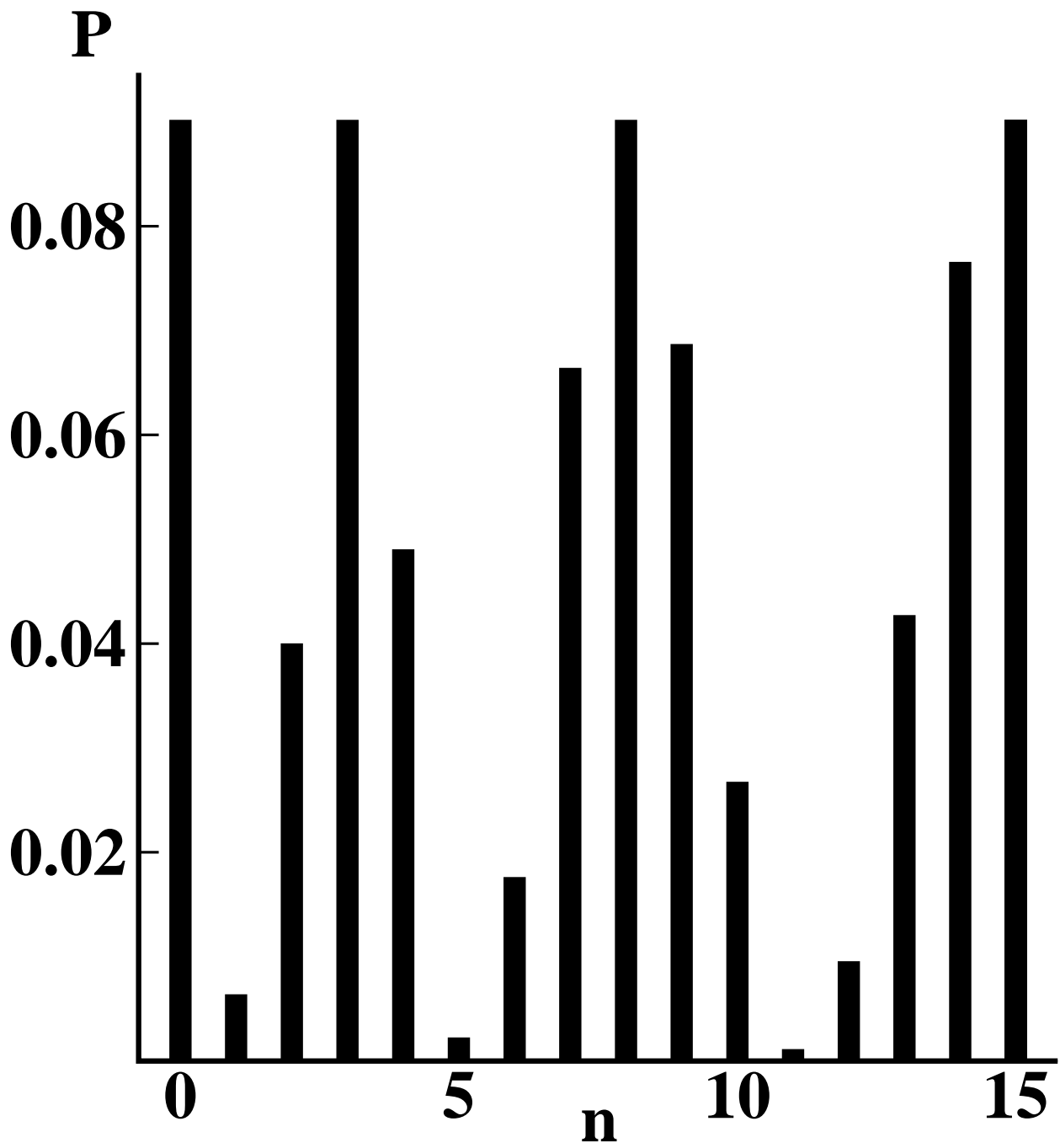
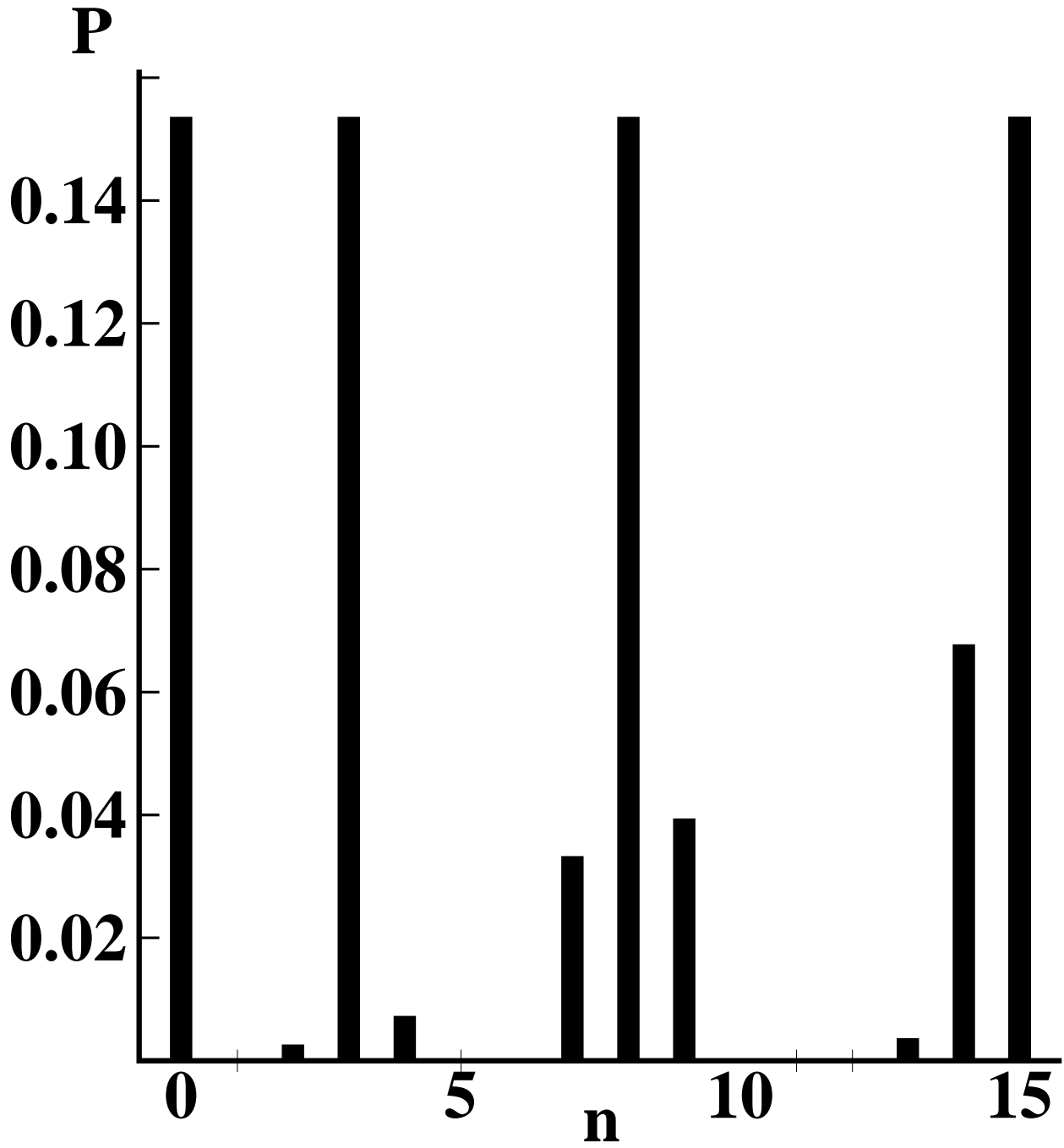
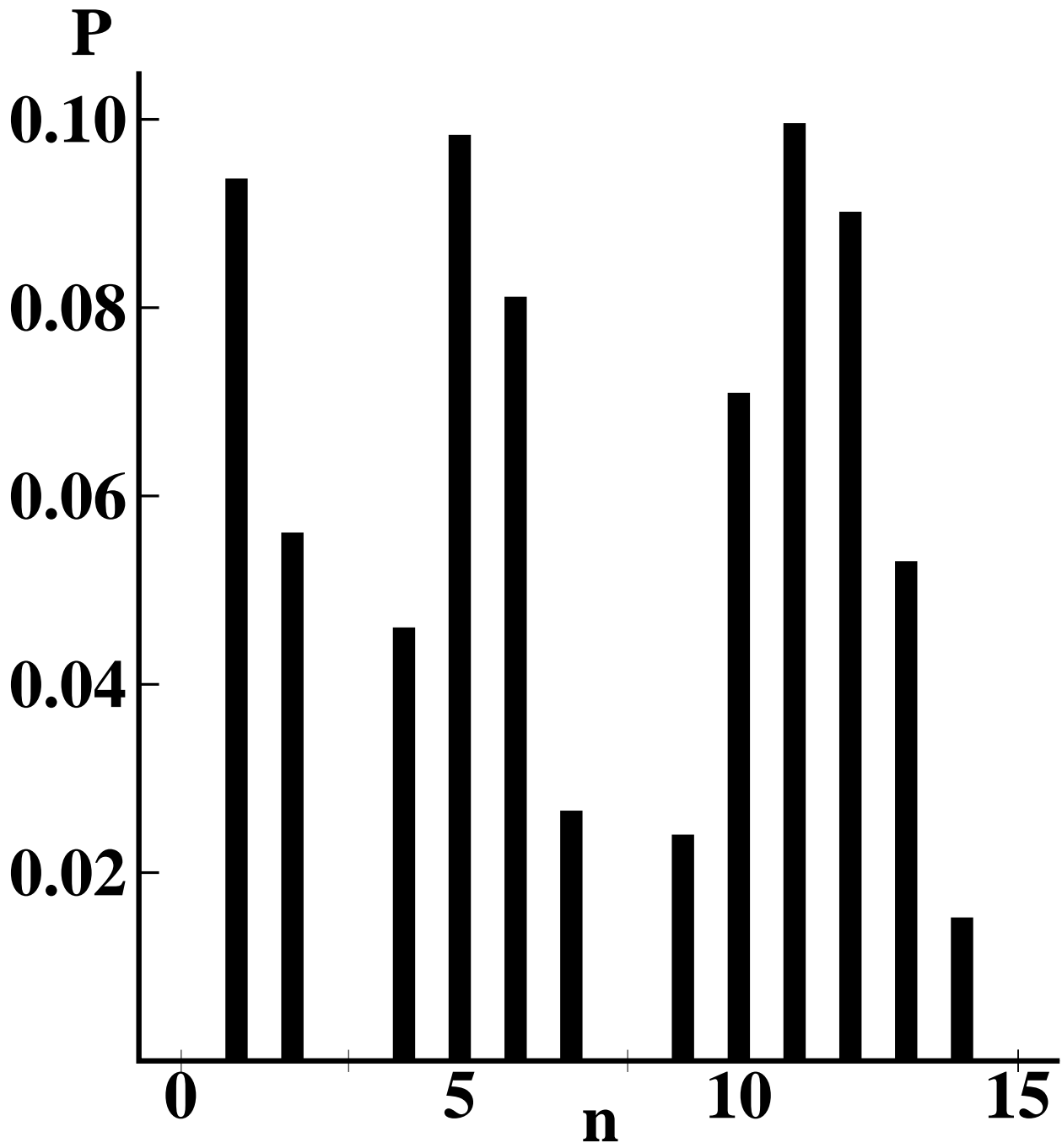
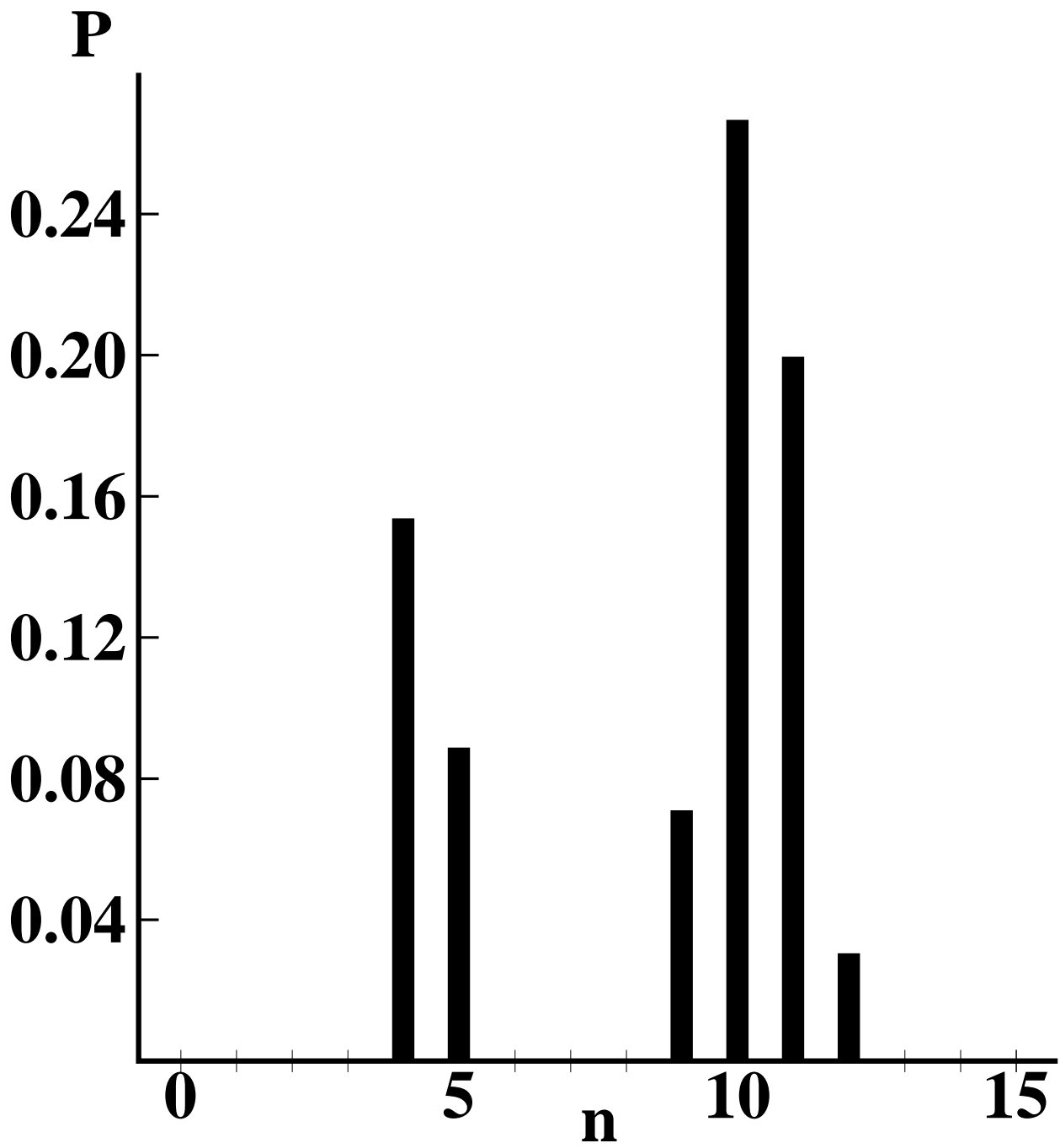
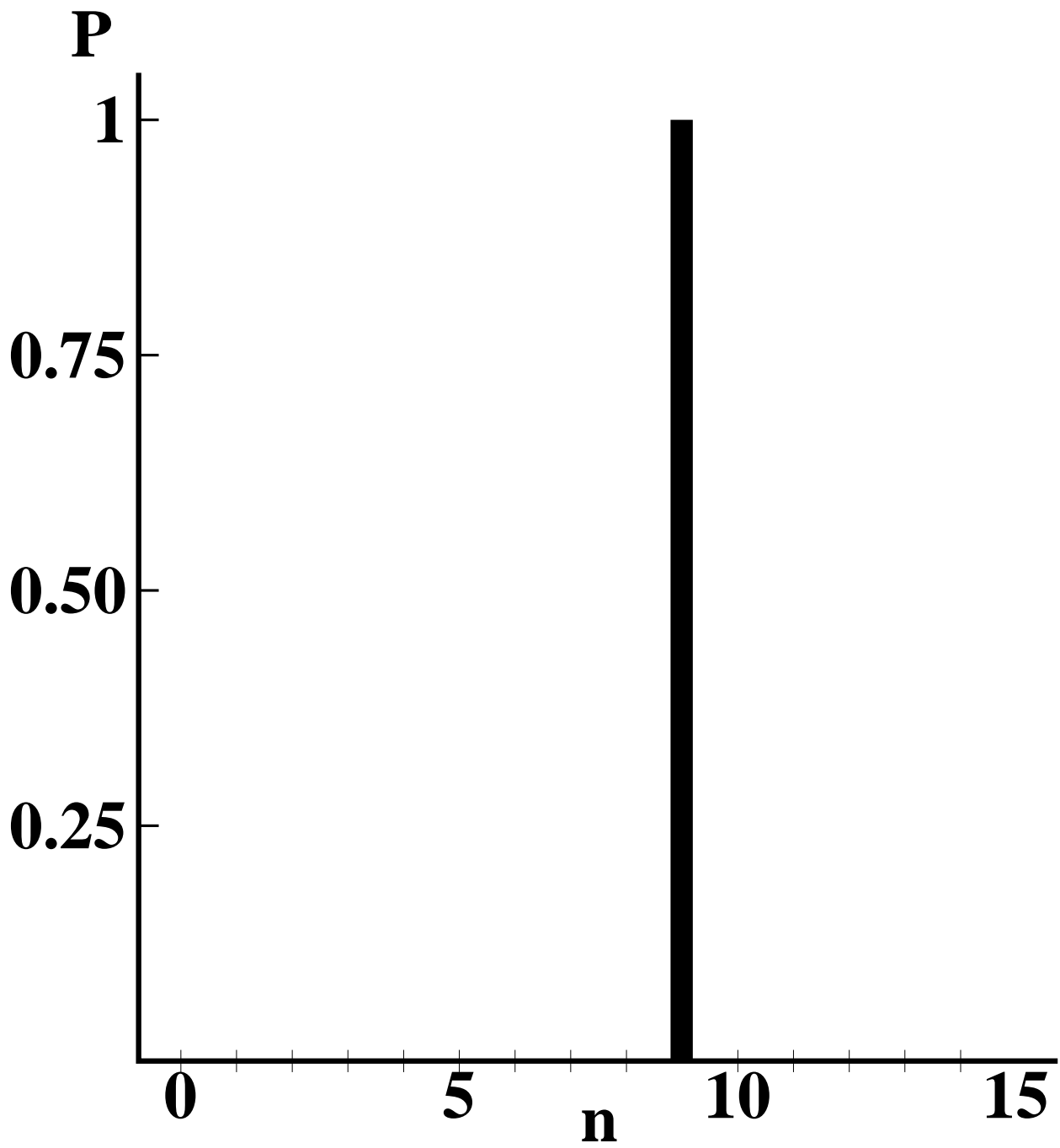
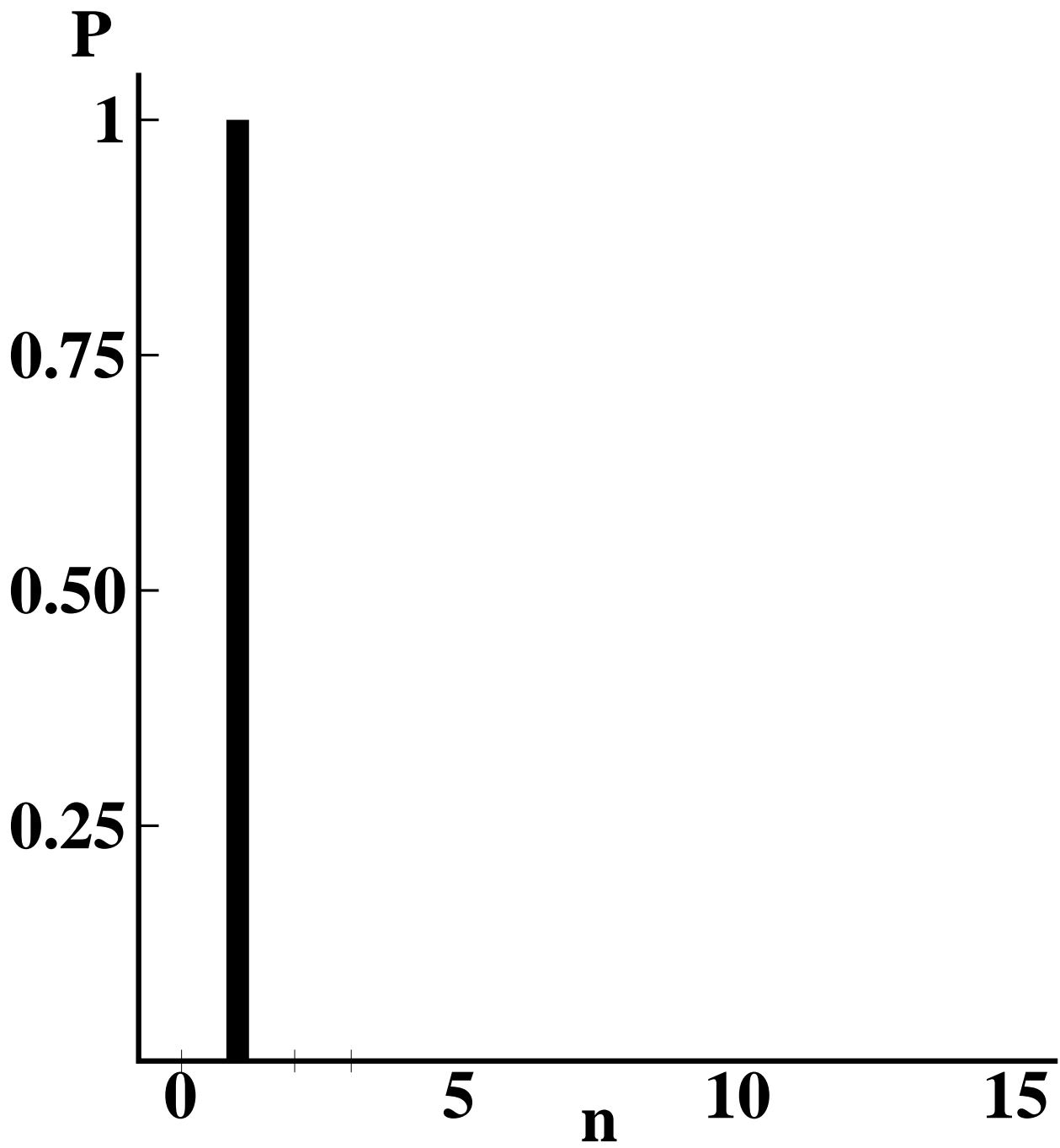
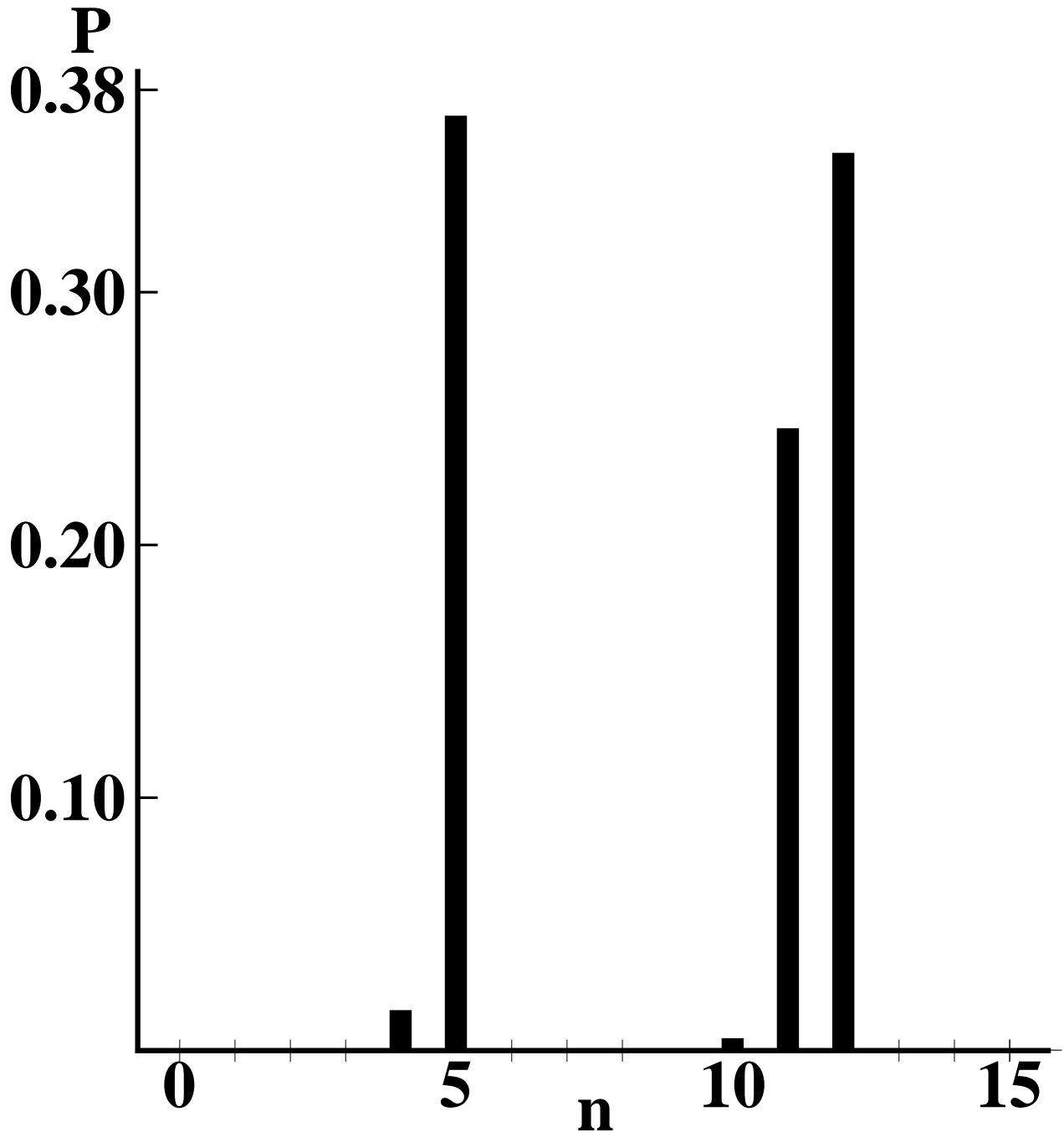
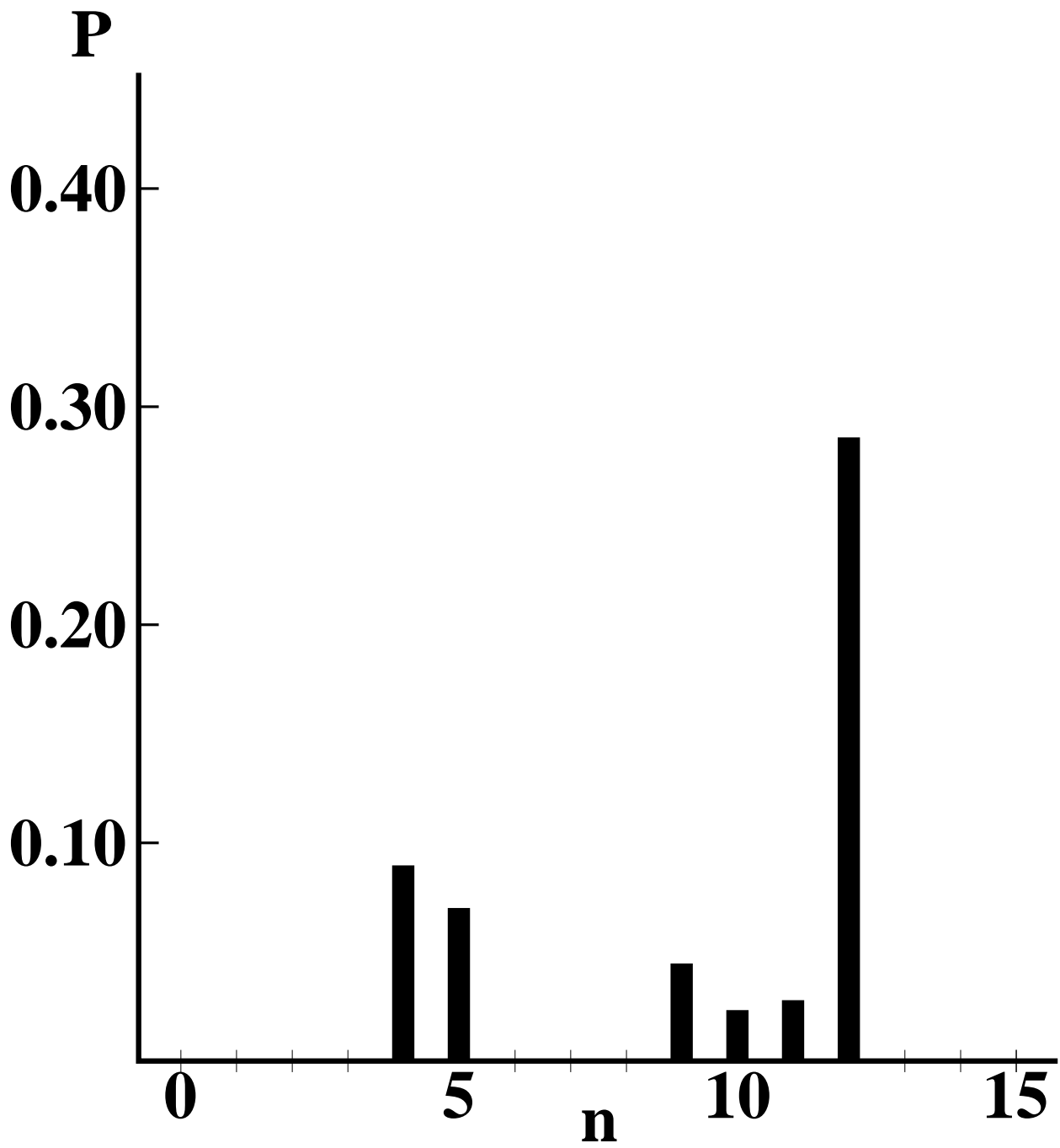
\end{document}